\documentclass[review]{elsarticle}

\usepackage{amssymb, tikz, float}
\usepackage{verbatim}
\usepackage{ifthen}

\usepackage{lineno,hyperref}

\journal{a journal}

\newtheorem{thm}{Theorem}

\newdefinition{df}{Definition}
\newdefinition{rmk}{Remark}

\newproof{pf}{Proof of Theorem \ref{thm_main}}

\begin{document}

\begin{frontmatter}

\title{\textsc{Atropos}-$k$ is \textsf{PSPACE}-complete\tnoteref{t1}}
\tnotetext[t1]{This work was partially supported by the Research Fund of Guangdong University of Foreign Studies (Nos. 297-ZW200011 and 297-ZW230018), and the National Natural Science Foundation of China (No. 61976104).}



\author[gdufs]{Chao Yang}
\ead{sokoban2007@163.com, yangchao@gdufs.edu.cn}

\author[gdmc]{Zhujun Zhang}
\ead{zhangzhujun1988@163.com}


\address[gdufs]{School of Mathematics and Statistics, Guangdong University of Foreign Studies, Guangzhou, 510006, China}

\address[gdmc]{Government Data Management Center of Fengxian District, Shanghai, 201499, China}

\begin{abstract}
Burke and Teng introduced a two-player combinatorial game \textsc{Atropos} based on Sperner's lemma, and showed that deciding whether one has a winning strategy for \textsc{Atropos} is \textsf{PSPACE}-complete. In the original \textsc{Atropos} game, the players must color a node adjacent to the last colored node. Burke and Teng also mentioned a variant \textsc{Atropos}-$k$ in which each move is at most of distance $k$ of the previous move, and asked a question on determining the computational complexity of this variant. In this paper, we answer this question by showing that for any fixed integer $k\ (k\geq 2)$, \textsc{Atropos}-$k$ is \textsf{PSPACE}-complete by reduction from True Quantified Boolean Formula (\textsc{TQBF}).   
\end{abstract}


\end{frontmatter}


\section{Introduction}

One of the major problems in studying combinatorial games is to determine their computational complexity. Combinatorial games old and new such as Chess \cite{fraenkel1981computing}, Checkers \cite{robson1984checker}, Go \cite{ls80,robson1983go}, Gobang \cite{reisch1980gobang} and Hex \cite{reisch1981hex}, were proved to be \textsf{PSPACE}-complete or \textsf{EXPTIME}-complete. Comprehensive surveys about the computational complexity of games can be found in \cite{hearn2009games,siegel2013combinatorial}.

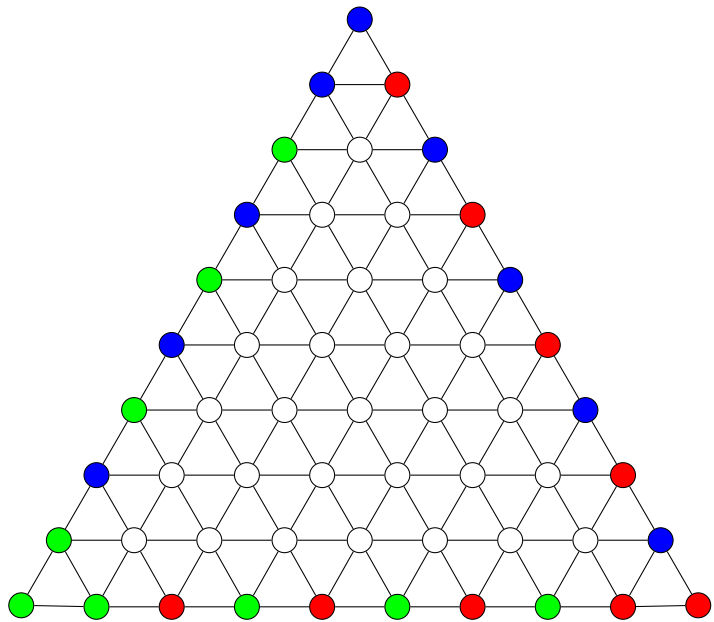
\begin{figure}[H]
\begin{center}
\begin{tikzpicture}


\foreach \x in {0,...,7} 
{ 
\ifthenelse{\x=1 \OR \x=3 \OR \x=5 \OR \x=7}
{\node (a\x) at (\x-0.5,-0.886) [circle,draw,fill=red] {}}
{\node (a\x) at (\x-0.5,-0.886) [circle,draw,fill=green] {}};
}

\foreach \x in {0,...,7} 
{ 
\ifthenelse{\x=1 \OR \x=3 \OR \x=5 \OR \x=7}
{\node (b\x) at (0.5*\x-1, 0.866*\x) [circle,draw,fill=blue] {}}
{\node (b\x) at (0.5*\x-1, 0.866*\x) [circle,draw,fill=green] {}};
}

\foreach \x in {0,...,7} 
{ 
\ifthenelse{\x=1 \OR \x=3 \OR \x=5 \OR \x=7}
{\node (c\x) at (7-0.5*\x, 0.866*\x) [circle,draw,fill=red] {}}
{\node (c\x) at (7-0.5*\x, 0.866*\x) [circle,draw,fill=blue] {}};
}

\node (A) at (3,0.866*8) [circle,draw, fill=blue] {};
\draw (c7)--(A)--(b7);
\node (C) at (-1.5, -0.866) [circle,draw, fill=green] {};
\draw (b0)--(C)--(a0);
\node (B) at (7.5, -0.866) [circle,draw, fill=red] {};
\draw (c0)--(B)--(a7);


\foreach \x in {0,...,6} 
{ 
\node (v0\x) at (\x,0) [circle,draw] {};
}

\foreach \x in {0,...,5} 
{ 
\node (v1\x) at (0.5+\x,0.866) [circle,draw] {};
}

\foreach \x in {0,...,4} 
{ 
\node (v2\x) at (1+\x,0.866*2) [circle,draw] {};
}

\foreach \x in {0,...,3} 
{ 
\node (v3\x) at (1.5+\x,0.866*3) [circle,draw] {};
}

\foreach \x in {0,...,2} 
{ 
\node (v4\x) at (2+\x,0.866*4) [circle,draw] {};
}

\foreach \x in {0,1} 
{ 
\node (v5\x) at (2.5+\x,0.866*5) [circle,draw] {};
}
\node (v60) at (3,0.866*6) [circle,draw] {};

\draw (v00)--(v01)--(v02)--(v03)--(v04)--(v05)--(v06)--(v15)--(v05)--(v14)--(v04)--(v13)--(v03)--(v12)--(v02)--(v11)--(v01)--(v10)--(v00);
\draw (v10)--(v11)--(v12)--(v13)--(v14)--(v15)--(v24)--(v14)--(v23)--(v13)--(v22)--(v12)--(v21)--(v11)--(v20)--(v10);
\draw (v20)--(v21)--(v22)--(v23)--(v24)--(v33)--(v23)--(v32)--(v22)--(v31)--(v21)--(v30)--(v20);
\draw (v30)--(v31)--(v32)--(v33)--(v42)--(v32)--(v41)--(v31)--(v40)--(v30);
\draw (v40)--(v41)--(v42)--(v51)--(v41)--(v50)--(v40);
\draw (v50)--(v51)--(v60)--(v50);

\draw (a0)--(a1)--(a2)--(a3)--(a4)--(a5)--(a6)--(a7)--(v06)--(a6)--(v05)--(a5)--(v04)--(a4)--(v03)--(a3)--(v02)--(a2)--(v01)--(a1)--(v00)--(a0);
\draw (b0)--(b1)--(b2)--(b3)--(b4)--(b5)--(b6)--(b7)--(v60)--(b6)--(v50)--(b5)--(v40)--(b4)--(v30)--(b3)--(v20)--(b2)--(v10)--(b1)--(v00)--(b0)--(a0);
\draw (c0)--(c1)--(c2)--(c3)--(c4)--(c5)--(c6)--(c7)--(v60)--(c6)--(v51)--(c5)--(v42)--(c4)--(v33)--(c3)--(v24)--(c2)--(v15)--(c1)--(v06)--(c0)--(a7);
\draw (b7)--(c7);

\end{tikzpicture}
\end{center}
\caption{An \textsc{Atropos} game board of size 7.}\label{fig_atropos}
\end{figure}

\textsc{Atropos} is a two-player combinatorial game with perfect information introduced by Burke and Teng in \cite{bt08}. The game is played on a triangular region of a triangular lattice (see Figure \ref{fig_atropos} for a game board of size $7$). The nodes of the outer boundary of the game board are precolored in the following way. The three vertices of the triangular game board are colored red, green, and blue respectively. Each node on the three outer edges of the triangular game board must be colored in one of the two colors of the two vertices on the same edge as that node. All the inner nodes of the game board are initially uncolored, the two players take turns to color the nodes in one of the three colors: red, green, and blue. A node can be colored only once. The first player, called \textit{hero} (the other player is called \textit{adversary}, following the convention of Burke and Teng), colors an arbitrary node of the game board at his first move. All subsequent moves should color a node that is adjacent the the node just colored by the other player in the previous move, if such an uncolored node exists. Otherwise, if all of the $6$ neighbors of the last-colored node have already been colored, the current player can choose to color an arbitrary uncolored node on the game board. If a player creates a \textit{rainbow triangle} (i.e. a triangle that receives three different colors in its three nodes), she or he loses the game immediately. As a consequence of Sperner's Lemma, either the hero or the adversary will win the game, there is no draw. A \textit{legal state} of \textsc{Atropos} is a partially colored game board with no rainbow triangles, and with a node specified as the last colored node. Burke and Teng defined the following decision problem associated with \textsc{Atropos} and determined its computational complexity.

\begin{df}[{\normalfont\scshape Atropos},\cite{bt08}]
    Given a legal \textsc{Atropos} state, determine whether the current player has a winning strategy.
\end{df}

\begin{thm}[\cite{bt08}]
    \textsc{Atropos} is \textsf{PSPACE}-complete.
\end{thm}

Burke and Teng also suggested studying the computational complexity of a variant of \textsc{Atropos}. In the original \textsc{Atropos}, the players should color a node adjacent to the last-colored node. This restriction can be relaxed by allowing the players to color a node that is of at most distance $k$ with the last-colored node (color an arbitrary node if there are no such nodes). We call this variant \textsc{Atropos}-$k$. Note that \textsc{Atropos}-$1$ is the same as the original \textsc{Atropos}, and \textsc{Atropos}-$\infty$ is identical to the \textsc{Unrestricted Atropos} defined by Burke and Teng in which the players could always choose to color any uncolored node.

\begin{df}[\textbf {\normalfont\scshape Atropos}-$k$]
    Let $k$ be a fixed positive integer or $k=\infty$. Given a legal \textsc{Atropos}-$k$ state, determine whether the current player has a winning strategy.
\end{df}

The main contribution of this paper is the following theorem, which answers a question posted by Burke and Teng \cite{bt08}.

\begin{thm}\label{thm_main}
    For any fixed integer $k\geq 2$, \textsc{Atropos}-$k$ is \textsf{PSPACE}-complete.
\end{thm}

We will prove Theorem \ref{thm_main} by reduction from a \textsf{PSPACE}-complete problem True Quantified Boolean Formula (\textsc{Tqbf}). 

\begin{df}[\textbf {\normalfont\scshape Tqbf},\cite{s12}]
    Given a fully quantified Boolean formula $\phi=Q_1 x_1 Q_2 x_2 \cdots Q_n x_n [\psi]$, where $Q_i=\forall$ or $Q_i=\exists$ ($i=1,\dots, n$) and $\psi=c_1 \wedge c_2 \wedge \dots \wedge c_m$ is in conjunctive normal form with $m$ clauses, determine whether the formula is true.
\end{df}

\textsc{Tqbf} has been used to show the \textsf{PSPACE}-hardness of several two-player combinatorial games including Generalized Geography, Go, Hex, and Othello  \cite{ls80,reisch1981hex,s12,iwata1994othello,cracsmaru2001ladders,zhang19}. \textsc{Atropos}-$1$ is shown to be \textsf{PSPACE}-complete also by reduction from \textsc{Tqbf} in \cite{bt08}. In their proof, a single path is used to simulate the process of assigning true or false to each Boolean variable. To settle the computational complexity of the general \textsc{Atropos}-$k$ ($k\geq 2$), we will use two paths for the simulation of the assignment of true or false, respectively.

The rest of the paper is organized as follows. Section \ref{sec_proof} gives the proof of Theorem \ref{thm_main}. Section \ref{sec_con} concludes with remarks on future work.

\section{Proof of Main Result}\label{sec_proof}
\newpage

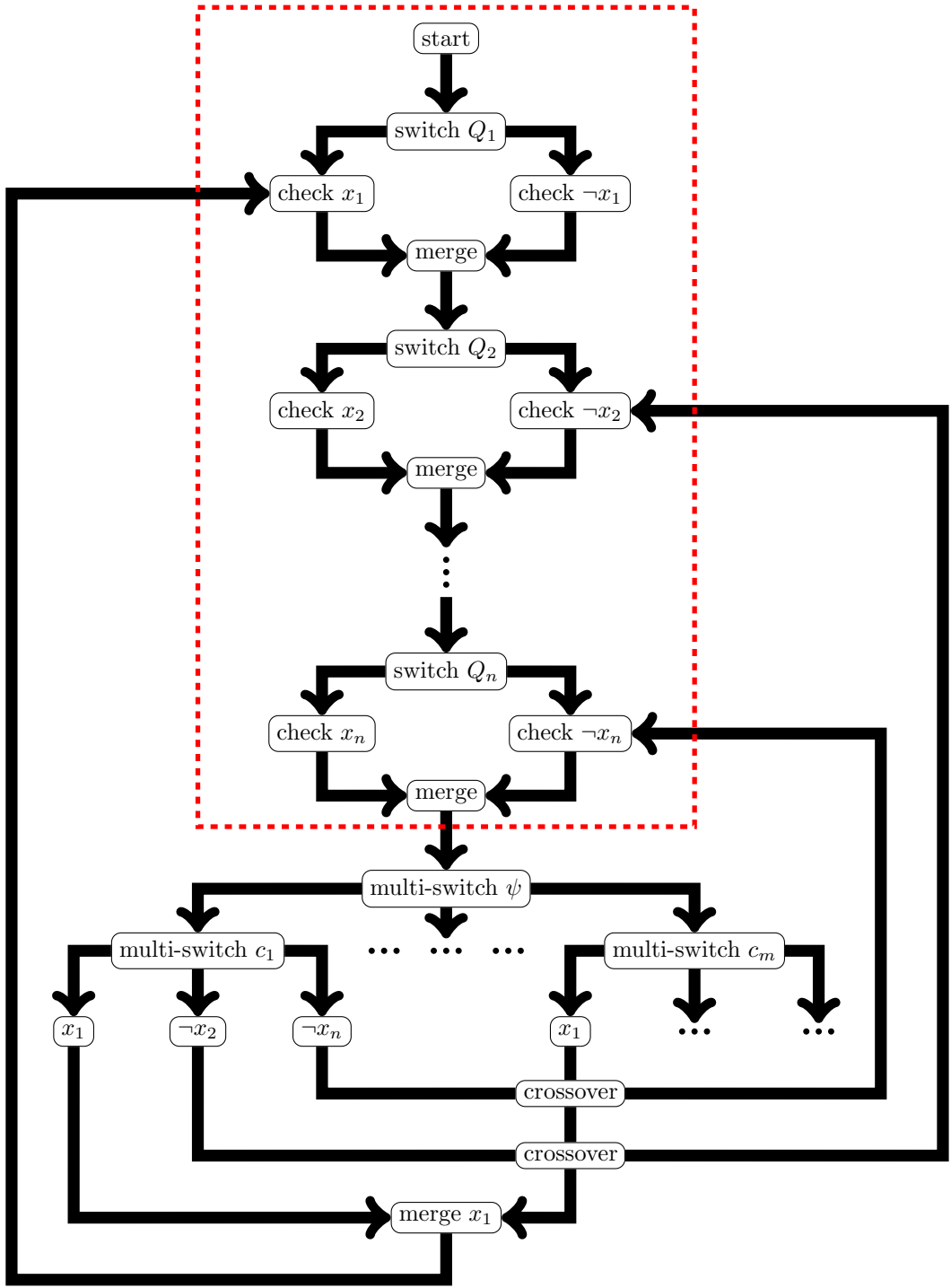
\begin{figure}[H]
\begin{center}
\begin{tikzpicture}[scale=0.92]

\node (start) at (2,0) [rectangle, rounded corners, draw] {start};
\node (s1) at (2,-1.5) [rectangle, rounded corners, draw] {switch $Q_1$};
\node (s2) at (2,-5) [rectangle, rounded corners, draw] {switch $Q_2$};
\node (s3) at (2,-10.2) [rectangle, rounded corners, draw] {switch $Q_n$};

\node (msc) at (2,-13.7) [rectangle, rounded corners, draw] {multi-switch $\psi$};
\node (msc1) at (-2,-14.7) [rectangle, rounded corners, draw] {multi-switch $c_1$};
\node (msc2) at (6,-14.7) [rectangle, rounded corners, draw] {multi-switch $c_m$};

\node (x1t) at (0,-2.5) [rectangle, rounded corners, draw] {check $x_1$};
\node (x1f) at (4,-2.5) [rectangle, rounded corners, draw] {check $\lnot x_1$};

\node (x2t) at (0,-6) [rectangle, rounded corners, draw] {check $x_2$};
\node (x2f) at (4,-6) [rectangle, rounded corners, draw] {check $\lnot x_2$};

\node (x3t) at (0,-11.2) [rectangle, rounded corners, draw] {check $x_n$};
\node (x3f) at (4,-11.2) [rectangle, rounded corners, draw] {check $\lnot x_n$};

\node (m1) at (2,-3.5) [rectangle, rounded corners, draw] {merge};
\node (m2) at (2,-7) [rectangle, rounded corners, draw] {merge};
\node (m3) at (2,-12.2) [rectangle, rounded corners, draw] {merge};

\draw [->, line width=5] (s1)--(0,-1.5)--(x1t);
\draw [->, line width=5] (s1)--(4,-1.5)--(x1f);
\draw [->, line width=5] (x1t)--(0,-3.5)--(m1);
\draw [->, line width=5] (x1f)--(4,-3.5)--(m1);

\draw [->, line width=5] (s2)--(0,-5)--(x2t);
\draw [->, line width=5] (s2)--(4,-5)--(x2f);
\draw [->, line width=5] (x2t)--(0,-7)--(m2);
\draw [->, line width=5] (x2f)--(4,-7)--(m2);

\draw [->, line width=5] (s3)--(0,-10.2)--(x3t);
\draw [->, line width=5] (s3)--(4,-10.2)--(x3f);
\draw [->, line width=5] (x3t)--(0,-12.2)--(m3);
\draw [->, line width=5] (x3f)--(4,-12.2)--(m3);

\draw [->, line width=5] (msc)--(-2,-13.7)--(msc1);
\draw [->, line width=5] (msc)--(6,-13.7)--(msc2);
\draw [->, line width=5] (msc)--(2,-14.5);

\draw [->, line width=5] (start)--(s1);
\draw [->, line width=5] (m1)--(s2);
\draw [->, line width=5] (m2)--(2,-8.2);
\draw [->, line width=5] (2,-9)--(s3);
\draw [->, line width=5] (m3)--(msc);

\fill (2,-8.4) circle [ radius=0.05];
\fill (2,-8.6) circle [ radius=0.05];
\fill (2,-8.8) circle [ radius=0.05];

\fill (0.8,-14.7) circle [ radius=0.05];
\fill (1,-14.7) circle [ radius=0.05];
\fill (1.2,-14.7) circle [ radius=0.05];

\fill (1.8,-14.7) circle [ radius=0.05];
\fill (2,-14.7) circle [ radius=0.05];
\fill (2.2,-14.7) circle [ radius=0.05];

\fill (2.8,-14.7) circle [ radius=0.05];
\fill (3,-14.7) circle [ radius=0.05];
\fill (3.2,-14.7) circle [ radius=0.05];


\fill (5.8,-16) circle [ radius=0.05];
\fill (6,-16) circle [ radius=0.05];
\fill (6.2,-16) circle [ radius=0.05];

\fill (7.8,-16) circle [ radius=0.05];
\fill (8,-16) circle [ radius=0.05];
\fill (8.2,-16) circle [ radius=0.05];


\node (v11) at (-4,-16) [rectangle, rounded corners, draw] {$x_1$};
\node (v12) at (-2,-16) [rectangle, rounded corners, draw] {$\lnot x_2$};
\node (v13) at (0,-16) [rectangle, rounded corners, draw] {$\lnot x_n$};

\draw [->, line width=5] (msc1)--(-4,-14.7)--(v11);
\draw [->, line width=5] (msc1)--(v12);
\draw [->, line width=5] (msc1)--(0,-14.7)--(v13);

\node (v21) at (4,-16) [rectangle, rounded corners, draw] {$x_1$};
\node (v22) at (6,-16) [rectangle, rounded corners] {};
\node (v23) at (8,-16) [rectangle, rounded corners] {};

\draw [->, line width=5] (msc2)--(4,-14.7)--(v21);
\draw [->, line width=5] (msc2)--(v22);
\draw [->, line width=5] (msc2)--(8,-14.7)--(v23);

\node (xing1) at (4,-18) [rectangle, rounded corners, draw] {crossover};
\node (xing2) at (4,-17) [rectangle, rounded corners, draw] {crossover};


\node (mv1) at (2,-19) [rectangle, rounded corners, draw] {merge $x_1$};
\draw [->, line width=5] (v11)--(-4,-19)--(mv1);
\draw [->, line width=5] (v21)--(xing2)--(xing1)--(4,-19)--(mv1);
\draw [->, line width=5] (mv1)--(2,-20)--(-5,-20)--(-5,-2.5)--(x1t);

\draw [->, line width=5] (v12)--(-2,-18)--(xing1)--(10,-18)--(10,-6)--(x2f);
\draw [->, line width=5] (v13)--(0,-17)--(xing2)--(9,-17)--(9,-11.2)--(x3f);

\draw [dashed, color=red, line width=2] (-2,0.5)--(6,0.5)--(6,-12.7)--(-2,-12.7)--(-2,0.5);

\end{tikzpicture}
\end{center}
\caption{The overall structure.}\label{fig_structure}
\end{figure}

\begin{pf} \textsc{Atropos}-$k\ (k\geq 2)$ is obviously in \textsf{PSPACE} by the same argument as \cite{bt08} to show that \textsc{Atropos}-$1$ is in \textsf{PSPACE}. In the rest of the proof, we will show that \textsc{Atropos}-$k$ is \textsf{PSPACE}-hard. To this end, for each fully quantified Boolean formula $\phi$, we will construct a legal \textsc{Atropos}-$2$ state such that $\phi$ is true if and only if the current player has a winning strategy for the corresponding \textsc{Atropos}-$2$ state. The construction can be generalized to show the \textsf{PSPACE}-hardness of \textsc{Atropos}-$k$ ($k\geq 3$) with little modification.

Given a formula $\phi$ with $n$ variables and $m$ clauses, the overall structure of the legal state of \textsc{Atropos}-$2$ corresponding to $\phi$ is illustrated in Figure \ref{fig_structure}. The rounded-corner rectangles represent different \textit{functional gadgets}, and they are connected to form a complete \textsc{Atropos}-$2$ state by arrow lines. These arrow lines represent the \textit{path gadgets}. The functional gadgets include the start gadget, switch gadgets, multi-switch gadgets,  merge gadgets, multi-merge gadgets, check gadgets, and crossover gadgets.

The overall game play of this legal state consists of two parts. Without loss of generality, we assume the current player of the state is the hero. The first part takes place in the area enclosed by the red dashed lines illustrated in Figure \ref{fig_structure}. Starting from the start gadget at the top, the two players take turns to assign a truth value to each variable of the formula $\phi$. The second part takes place on the game board outside the red dashed lines, where the two players decide to check the truth value of a particular literal of a particular clause. The outcome will determine who is the winner. 

In what follows, we will explain the path gadget and all the functional gadgets in detail, and how to combine them to work as sketched above.


The path gadget is illustrated in Figure \ref{fig_path}. It is a chain of uncolored nodes surrounded by two layers of red nodes. The distance between consecutive empty nodes along the path is either $1$ or $2$. It is also arranged in a way such that for each uncolored node, there are exactly two other uncolored nodes within a distance of $2$. With these arrangements, it is easy to check that once a player enters a path gadget, each empty node must be colored and can be colored without creating a rainbow triangle by the two players alternatively. Thus path gadgets can lead the players from one functional gadget to another. The path gadget is very flexible, it can have arbitrary length and make turns, which guarantees that we can always connect two functional gadgets by a path gadget. The \textit{length} of a path gadget is the total number of empty nodes it contains.


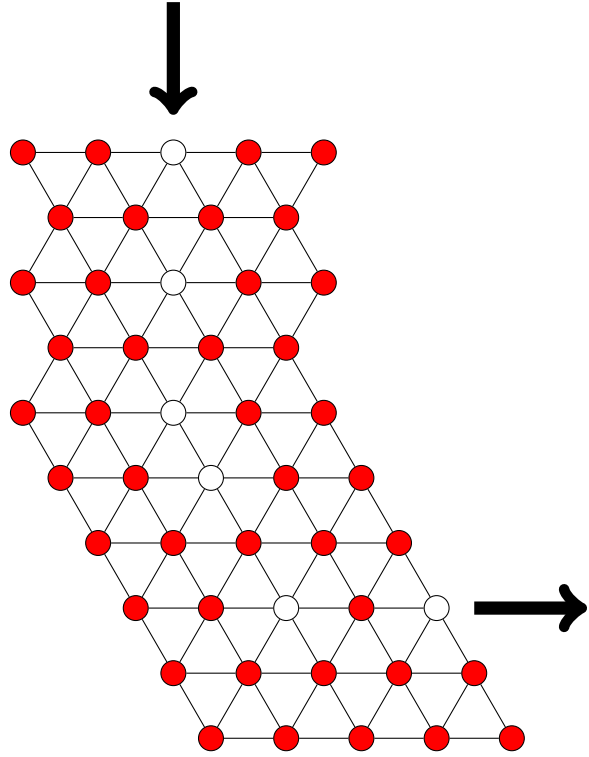
\begin{figure}[H]
\begin{center}
\begin{tikzpicture}

\draw [->, line width=5] (2,2)--(2,0.5);
\draw [->, line width=5] (6,-7*0.866)--(7.5,-7*0.866);

\foreach \x in {0,1,3,4} 
{ 
\node (v0\x) at (\x,0) [circle,draw, fill=red] {};
}
\foreach \x in {2} 
{ 
\node (v0\x) at (\x,0) [circle,draw] {};
}
\foreach \x in {0,1,2,3} 
{ 
\node (v1\x) at (0.5+\x,-0.866) [circle,draw, fill=red] {};
}
\foreach \x in {0,1,3,4} 
{ 
\node (v2\x) at (\x,-0.866*2) [circle,draw, fill=red] {};
}
\foreach \x in {2} 
{ 
\node (v2\x) at (\x,-0.866*2) [circle,draw] {};
}
\foreach \x in {0,1,2,3} 
{ 
\node (v3\x) at (0.5+\x,-0.866*3) [circle,draw, fill=red] {};
}

\foreach \x in {0,1,3,4} 
{ 
\node (v4\x) at (\x,-0.866*4) [circle,draw, fill=red] {};
}
\foreach \x in {2} 
{ 
\node (v4\x) at (\x,-0.866*4) [circle,draw] {};
}

\foreach \x in {0,1,3,4} 
{ 
\node (v5\x) at (0.5+\x,-0.866*5) [circle,draw, fill=red] {};
}
\foreach \x in {2} 
{ 
\node (v5\x) at (0.5+\x,-0.866*5) [circle,draw] {};
}

\foreach \x in {0,...,4} 
{ 
\node (v6\x) at (1+\x,-0.866*6) [circle,draw, fill=red] {};
}

\foreach \x in {0,1,3} 
{ 
\node (v7\x) at (1.5+\x,-0.866*7) [circle,draw, fill=red] {};
}
\foreach \x in {2,4} 
{ 
\node (v7\x) at (1.5+\x,-0.866*7) [circle,draw] {};
}

\foreach \x in {0,...,4} 
{ 
\node (v8\x) at (2+\x,-0.866*8) [circle,draw, fill=red] {};
}

\foreach \x in {0,...,4} 
{ 
\node (v9\x) at (2.5+\x,-0.866*9) [circle,draw, fill=red] {};
}

\draw (v00)--(v01)--(v02)--(v03)--(v04)--(v13)--(v03)--(v12)--(v02)--(v11)--(v01)--(v10)--(v00);
\draw (v10)--(v11)--(v12)--(v13);
\draw (v20)--(v21)--(v22)--(v23)--(v24)--(v13)--(v23)--(v12)--(v22)--(v11)--(v21)--(v10)--(v20);
\draw (v30)--(v31)--(v32)--(v33);
\draw (v20)--(v30)--(v21)--(v31)--(v22)--(v32)--(v23)--(v33)--(v24);

\draw (v40)--(v41)--(v42)--(v43)--(v44)--(v33)--(v43)--(v32)--(v42)--(v31)--(v41)--(v30)--(v40);
\draw (v50)--(v51)--(v52)--(v53)--(v54)--(v44)--(v53)--(v43)--(v52)--(v42)--(v51)--(v41)--(v50)--(v40);

\draw (v60)--(v61)--(v62)--(v63)--(v64)--(v54)--(v63)--(v53)--(v62)--(v52)--(v61)--(v51)--(v60)--(v50);

\draw (v70)--(v71)--(v72)--(v73)--(v74)--(v64)--(v73)--(v63)--(v72)--(v62)--(v71)--(v61)--(v70)--(v60);
\draw (v80)--(v81)--(v82)--(v83)--(v84)--(v74)--(v83)--(v73)--(v82)--(v72)--(v81)--(v71)--(v80)--(v70);
\draw (v90)--(v91)--(v92)--(v93)--(v94)--(v84)--(v93)--(v83)--(v92)--(v82)--(v91)--(v81)--(v90)--(v80);

\end{tikzpicture}
\end{center}
\caption{The path gadget.}\label{fig_path}
\end{figure}

Note that there is no inherent direction for the path gadgets, the direction is forced by the fact that the game is started from the start gadget which is illustrated in Figure \ref{fig_start}.

In Figure \ref{fig_start}, the node labelled $l$ is the last-colored node of this legal state, and there is exactly one empty node within distance $2$ to the node $l$. By the rules of \textsc{Atropos}-$2$, the current player, namely the hero, must color the node labelled $s$. This move kicks off the game, and the two players move on to the unique path gadget attaching to the start gadget. We consider the empty node $s$ and the surrounding red nodes to be the start gadget, and the empty node labelled $P$ in Figure \ref{fig_start} is considered to be the first empty node of the path gadget attaching to the start gadget. In this paper, we follow the convention that a node labelled by a lowercase letter belongs to a functional gadget, while a node labelled by an uppercase letter belongs to a path gadget.


\begin{figure}[H]
\begin{center}
\begin{tikzpicture}

\foreach \x in {0,...,2} 
{ 
\node (a\x) at (-0.5+\x,0.866) [circle,draw, fill=red] {};
}

\foreach \x in {-1,0,...,2} 
{ 
\node (v0\x) at (\x,0) [circle,draw, fill=red] {};
}

\foreach \x in {-1,0,2,3} 
{ 
\node (v1\x) at (-0.5+\x,-0.866) [circle,draw, fill=red] {};
}
\foreach \x in {1} 
{ 
\node (v1\x) at (-0.5+\x,-0.866) [circle,draw, fill=red] {$l$};
}

\foreach \x in {0,1,3,4} 
{ 
\node (v2\x) at (-1+\x,-0.866*2) [circle,draw, fill=red] {};
}
\foreach \x in {2} 
{ 
\node (v2\x) at (-1+\x,-0.866*2) [circle,draw] {$s$};
}

\foreach \x in {0,...,3} 
{ 
\node (v3\x) at (-0.5+\x,-0.866*3) [circle,draw, fill=red] {};
}
\foreach \x in {0,1,3,4} 
{ 
\node (v4\x) at (-1+\x,-0.866*4) [circle,draw, fill=red] {};
}
\foreach \x in {2} 
{ 
\node (v4\x) at (-1+\x,-0.866*4) [circle,draw] {$P$};
}

\draw (v00)--(v01)--(v02);
\draw (v10)--(v11)--(v12)--(v13)--(v02)--(v12)--(v01)--(v11)--(v00)--(v10);

\draw (v20)--(v21)--(v22)--(v23)--(v24)--(v13)--(v23)--(v12)--(v22)--(v11)--(v21)--(v10)--(v20);

\draw (v30)--(v31)--(v32)--(v33);
\draw (v40)--(v41)--(v42)--(v43)--(v44)--(v33)--(v43)--(v32)--(v42)--(v31)--(v41)--(v30)--(v40);
\draw (v24)--(v33)--(v23)--(v32)--(v22)--(v31)--(v21)--(v30)--(v20);

\draw (v02)--(a2)--(v01)--(a1)--(v00)--(a0)--(v0-1)--(v00);
\draw (a0)--(a1)--(a2);
\draw (v20)--(v1-1)--(v10)--(v0-1)--(v1-1);

\draw [->, line width=5] (1,-0.866*4-0.5)--(1,-0.866*4-2);

\end{tikzpicture}
\end{center}
\caption{The start gadget.}\label{fig_start}
\end{figure}
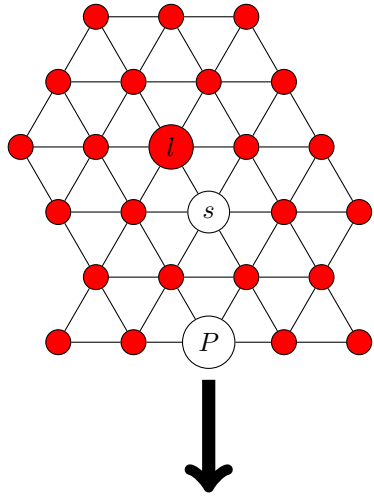

The switch gadget is illustrated in Figure \ref{fig_switch}. The two players enter the switch gadget from a path gadget on the top. When it is one player's turn to color the unique empty node $s$ of the switch gadget, then his opponent gets the chance to choose to color either the node $L$ or the node $R$, resulting in leaving the switch gadget and entering a path gadget either to the left or to the right. Note that the distance between $L$ and $R$ is 3, so once one of them was colored the other node could not be colored.


\begin{figure}[H]
\begin{center}
\begin{tikzpicture}

\draw [->, line width=5] (2,2)--(2,0.5);
\draw [->, line width=5] (4,-3*0.866)--(5.5,-3*0.866);
\draw [->, line width=5] (0,-3*0.866)--(-1.5,-3*0.866);

\foreach \x in {0,1,3,4} 
{ 
\node (v0\x) at (\x,0) [circle,draw, fill=red] {};
}
\foreach \x in {2} 
{ 
\node (v0\x) at (\x,0) [circle,draw] {$T$};
}

\foreach \x in {0,1,2,3} 
{ 
\node (v1\x) at (0.5+\x,-0.866) [circle,draw, fill=red] {};
}

\foreach \x in {0,1,3,4} 
{ 
\node (v2\x) at (\x,-0.866*2) [circle,draw, fill=red] {};
}
\foreach \x in {2} 
{ 
\node (v2\x) at (\x,-0.866*2) [circle,draw] {$s$};
}

\foreach \x in {1,2} 
{ 
\node (v3\x) at (0.5+\x,-0.866*3) [circle,draw, fill=red] {};
}
\foreach \x in {0} 
{ 
\node (v3\x) at (0.5+\x,-0.866*3) [circle,draw] {$L$};
}
\foreach \x in {3} 
{ 
\node (v3\x) at (0.5+\x,-0.866*3) [circle,draw] {$R$};
}

\foreach \x in {0,...,4} 
{ 
\node (v4\x) at (\x,-0.866*4) [circle,draw, fill=red] {};
}

\foreach \x in {0,...,3} 
{ 
\node (v5\x) at (0.5+\x,-0.866*5) [circle,draw, fill=red] {};
}

\draw (v00)--(v01)--(v02)--(v03)--(v04)--(v13)--(v03)--(v12)--(v02)--(v11)--(v01)--(v10)--(v00);
\draw (v10)--(v11)--(v12)--(v13);
\draw (v20)--(v21)--(v22)--(v23)--(v24)--(v13)--(v23)--(v12)--(v22)--(v11)--(v21)--(v10)--(v20);
\draw (v30)--(v31)--(v32)--(v33);
\draw (v20)--(v30)--(v21)--(v31)--(v22)--(v32)--(v23)--(v33)--(v24);

\draw (v40)--(v41)--(v42)--(v43)--(v44)--(v33)--(v43)--(v32)--(v42)--(v31)--(v41)--(v30)--(v40);
\draw (v50)--(v51)--(v52)--(v53);
\draw (v44)--(v53)--(v43)--(v52)--(v42)--(v51)--(v41)--(v50)--(v40);

\end{tikzpicture}
\end{center}
\caption{The switch gadget.}\label{fig_switch}
\end{figure}
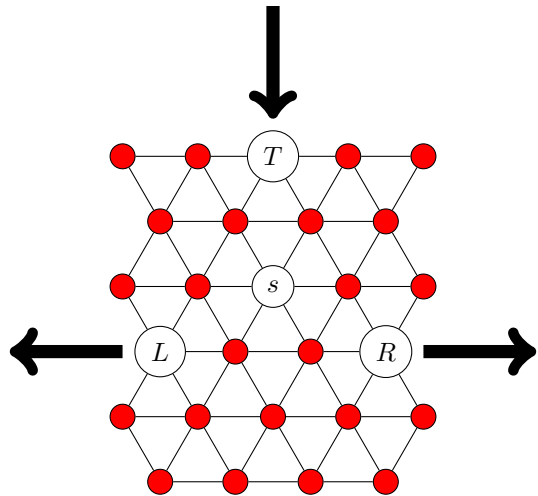

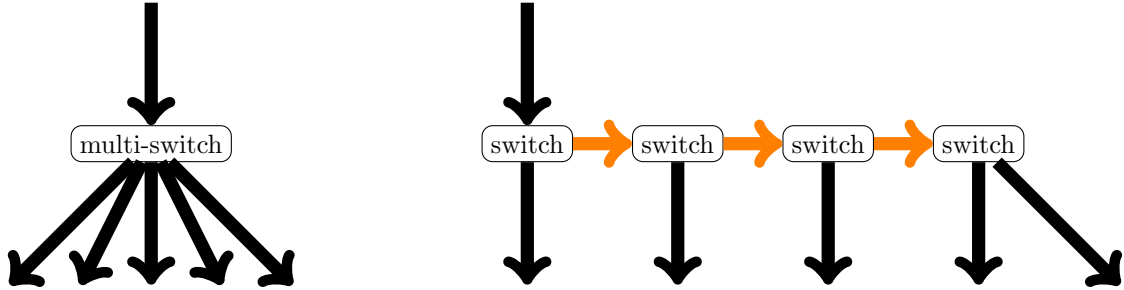
\begin{figure}[H]
\begin{center}
\begin{tikzpicture}

\node (i) at (0,2) {};
\node (a) at (0,0) [rectangle, rounded corners, draw] {multi-switch};
\node (o1) at (-2,-2) {};
\node (o2) at (-1,-2) {};
\node (o3) at (0,-2) {};
\node (o4) at (1,-2) {};
\node (o5) at (2,-2) {};

\draw [<-, line width=5] (a)--(i);
\draw [->, line width=5] (a)--(o1);
\draw [->, line width=5] (a)--(o2);
\draw [->, line width=5] (a)--(o3);
\draw [->, line width=5] (a)--(o4);
\draw [->, line width=5] (a)--(o5);

\node (b1) at (5,0) [rectangle, rounded corners, draw] {switch};
\node (b2) at (7,0) [rectangle, rounded corners, draw] {switch};
\node (b3) at (9,0) [rectangle, rounded corners, draw] {switch};
\node (b4) at (11,0) [rectangle, rounded corners, draw] {switch};

\node (bi) at (5,2) {};
\node (bo1) at (5,-2) {};
\node (bo2) at (7,-2) {};
\node (bo3) at (9,-2) {};
\node (bo4) at (11,-2) {};
\node (bo5) at (13,-2) {};

\draw [->, line width=5, color=orange] (b1)--(b2);
\draw [->, line width=5, color=orange] (b2)--(b3);
\draw [->, line width=5, color=orange] (b3)--(b4);

\draw [->, line width=5] (bi)--(b1);
\draw [->, line width=5] (b1)--(bo1);
\draw [->, line width=5] (b2)--(bo2);
\draw [->, line width=5] (b3)--(bo3);
\draw [->, line width=5] (b4)--(bo4);
\draw [->, line width=5] (b4)--(bo5);

\end{tikzpicture}
\end{center}
\caption{The multi-switch gadget (left) implemented by several switch gadgets (right).}\label{fig_multi_switch}
\end{figure}

Besides the standard switch gadget just introduced above, we also need multi-switch gadgets which have more than two out-going paths. This can be done by connecting several switch gadgets together as illustrated in Figure \ref{fig_multi_switch}. On the left of Figure \ref{fig_multi_switch}, there is a symbolical representation of a multi-switch gadget with $5$ out-going paths, which is in fact implemented by a group of $4$ standard switch gadgets as shown on the right of Figure \ref{fig_multi_switch}. Note that the lengths of the three orange edges gadgets are odd. Recall that the length is the number of empty nodes in the path gadgets. So it is the same player to choose the out-going path for all the $4$ standard switch gadgets, which is equivalent to a multi-switch gadget in that the same player is responsible for choosing one of many out-going paths.


\begin{figure}[H]
\begin{center}
\begin{tikzpicture}

\draw [->, line width=5] (3,-2.5*0.9)--(3,-2.5*0.9-1.5);
\draw [<-, line width=5] (5.5,0)--(7,0);
\draw [<-, line width=5] (-0.5,0)--(-2,0);

\foreach \x in {0} 
{ 
\node (v0\x) at (\x,0) [circle,draw] {$L$};
}
\foreach \x in {5} 
{ 
\node (v0\x) at (\x,0) [circle,draw] {$R$};
}
\foreach \x in {2} 
{ 
\node (v0\x) at (\x,0) [circle,draw] {$a$};
}
\foreach \x in {3} 
{ 
\node (v0\x) at (\x,0) [circle,draw] {$b$};
}
\foreach \x in {1} 
{ 
\node (v0\x) at (\x,0) [circle,draw,fill=red] {};
}
\foreach \x in {4} 
{ 
\node (v0\x) at (\x,0) [circle,draw,fill=green] {};
}

\foreach \x in {1,2,3} 
{ 
\node (v1\x) at (0.5+\x,0.866) [circle,draw,fill=blue] {};
}
\foreach \x in {0,-1,5} 
{ 
\node (v1\x) at (0.5+\x,0.866) [circle,draw,fill=red] {};
}
\foreach \x in {4} 
{ 
\node (v1\x) at (0.5+\x,0.866) [circle,draw,fill=green] {};
}

\foreach \x in {0,1,2,3} 
{ 
\node (v2\x) at (1+\x,0.866*2) [circle,draw,fill=blue] {};
}
\foreach \x in {4} 
{ 
\node (v2\x) at (1+\x,0.866*2) [circle,draw,fill=green] {};
}
\foreach \x in {-2,-1,5} 
{ 
\node (v2\x) at (1+\x,0.866*2) [circle,draw,fill=red] {};
}

\foreach \x in {1,3,4,5,6,0} 
{ 
\node (a\x) at (\x-0.5,-0.886) [circle,draw,fill=red] {};
}
\foreach \x in {2} 
{ 
\node (a\x) at (\x-0.5,-0.886) [circle,draw,fill=green] {};
}

\foreach \x in {1,2,4,5,6,0,-1} 
{ 
\node (b\x) at (\x,-0.886*2) [circle,draw,fill=red] {};
}
\foreach \x in {3} 
{ 
\node (b\x) at (\x,-0.886*2) [circle,draw] {$M$};
}

\draw (v00)--(v01)--(v02)--(v03)--(v04)--(v05)--(v14)--(v04)--(v13)--(v03)--(v12)--(v02)--(v11)--(v01)--(v10)--(v00);
\draw (v10)--(v11)--(v12)--(v13)--(v14);
\draw (v20)--(v11)--(v21)--(v12)--(v22)--(v13)--(v23);
\draw (v10)--(v20)--(v21)--(v22)--(v23)--(v14);

\draw (a1)--(a2)--(a3)--(a4)--(a5)--(v05);
\draw (v00)--(a1)--(v01)--(a2)--(v02)--(a3)--(v03)--(a4)--(v04)--(a5);

\draw (a1)--(b1)--(a2)--(b2)--(a3)--(b3)--(a4)--(b4)--(a5);
\draw (b1)--(b2)--(b3)--(b4);

\draw (v00)--(v1-1)--(v2-2)--(v2-1)--(v1-1)--(v10)--(v2-1)--(v20);
\draw (v05)--(v15)--(v25)--(v24)--(v15)--(v14)--(v24)--(v23);

\draw (v05)--(a6)--(b6)--(b5)--(a6)--(a5)--(b5)--(b4);
\draw (v00)--(a0)--(b-1)--(b0)--(a0)--(a1)--(b0)--(b1);

\end{tikzpicture}
\end{center}
\caption{The merge gadget.}\label{fig_merge}
\end{figure}
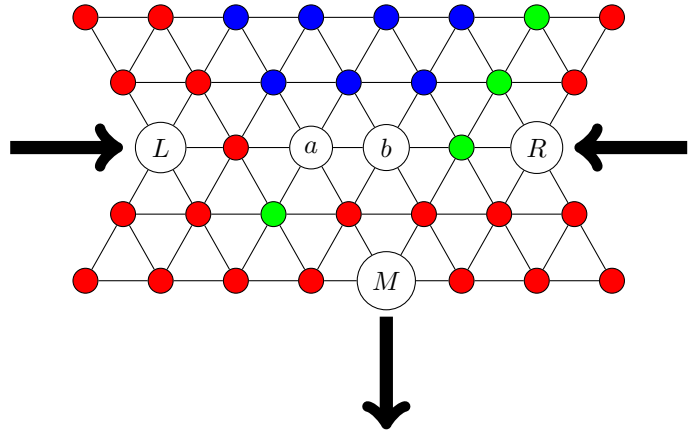

The merge gadget combines two paths into one. As illustrated in Figure \ref{fig_merge}, if the players come from the left, the node labelled $L$ is colored first, then the node $a$. To avoid getting a rainbow triangle, the node $a$ can only be colored red. After that, node $b$ cannot be colored unless creating a rainbow triangle. In order to avoid losing immediately, the next player has to color node $M$ and leave the gadget, since the distance between node $a$ and $R$ is 3. Similarly, if the players come from the right, then the nodes $R$, $b$, and $M$ are colored in sequence. Note that the distance between node $M$ and $L,R$ is at least 3. Therefore, no matter whether players come from the left or the right, the only way out is the path gadget at the bottom.

We also need multi-merge gadgets in our reduction. Because in the second part of the game play, we may need to combine the paths from more than two of the same literal ($x_i$ or $\lnot x_i$) into one path gadget before heading to the check gadget for that literal. We can build a multi-merge gadget by using a group of merge gadgets similar to the method we used to construct a multi-switch gadget.


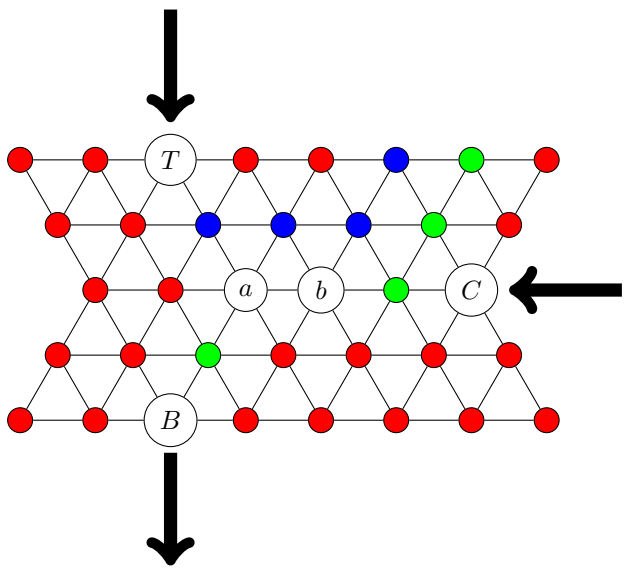
\begin{figure}[H]
\begin{center}
\begin{tikzpicture}

\draw [->, line width=5] (2,2)--(2,0.5);
\draw [->, line width=5] (2,-4.5*0.866)--(2,-4.5*0.866-1.5);
\draw [->, line width=5] (8,-2*0.866)--(6.5,-2*0.866);

\foreach \x in {0,1,3,4,7} 
{ 
\node (v0\x) at (\x,0) [circle,draw,fill=red] {};
}
\foreach \x in {2} 
{ 
\node (v0\x) at (\x,0) [circle,draw] {$T$};
}
\foreach \x in {5} 
{ 
\node (v0\x) at (\x,0) [circle,draw,fill=blue] {};
}
\foreach \x in {6} 
{ 
\node (v0\x) at (\x,0) [circle,draw,fill=green] {};
}

\foreach \x in {0,1,6} 
{ 
\node (v1\x) at (0.5+\x,-0.866) [circle,draw,fill=red] {};
}
\foreach \x in {2,3,4} 
{ 
\node (v1\x) at (0.5+\x,-0.866) [circle,draw,fill=blue] {};
}
\foreach \x in {5} 
{ 
\node (v1\x) at (0.5+\x,-0.866) [circle,draw,fill=green] {};
}

\foreach \x in {2} 
{ 
\node (v2\x) at (1+\x,-0.866*2) [circle,draw] {$a$};
}
\foreach \x in {3} 
{ 
\node (v2\x) at (1+\x,-0.866*2) [circle,draw] {$b$};
}
\foreach \x in {5} 
{ 
\node (v2\x) at (1+\x,-0.866*2) [circle,draw] {$C$};
}
\foreach \x in {0,1} 
{ 
\node (v2\x) at (1+\x,-0.866*2) [circle,draw,fill=red] {};
}
\foreach \x in {4} 
{ 
\node (v2\x) at (1+\x,-0.866*2) [circle,draw,fill=green] {};
}

\foreach \x in {0,1,3,4,5,6} 
{ 
\node (v3\x) at (0.5+\x,-0.866*3) [circle,draw,fill=red] {};
}
\foreach \x in {2} 
{ 
\node (v3\x) at (0.5+\x,-0.866*3) [circle,draw,fill=green] {};
}

\foreach \x in {0,1,3,4,5,6,7} 
{ 
\node (v4\x) at (\x,-0.866*4) [circle,draw,fill=red] {};
}
\foreach \x in {2} 
{ 
\node (v4\x) at (\x,-0.866*4) [circle,draw] {$B$};
}

\draw (v00)--(v01)--(v02)--(v03)--(v04)--(v05)--(v06)--(v07)--(v16)--(v06)--(v15)--(v05)--(v14);
\draw (v10)--(v11)--(v12)--(v13)--(v14)--(v15)--(v16)--(v25)--(v15)--(v24)--(v14)--(v23)--(v13)--(v22)--(v12)--(v21)--(v11)--(v20)--(v10);
\draw (v00)--(v10)--(v01)--(v11)--(v02)--(v12)--(v03)--(v13)--(v04)--(v14);

\draw (v20)--(v21)--(v22)--(v23)--(v24)--(v25);
\draw (v30)--(v31)--(v32)--(v33)--(v34)--(v35)--(v36)--(v25)--(v35)--(v24)--(v34)--(v23)--(v33)--(v22)--(v32)--(v21)--(v31)--(v20)--(v30);

\draw (v40)--(v41)--(v42)--(v43)--(v44)--(v33)--(v43)--(v32)--(v42)--(v31)--(v41)--(v30)--(v40);
\draw (v44)--(v45)--(v46)--(v35)--(v45)--(v34)--(v44);
\draw (v46)--(v36)--(v47)--(v46);

\end{tikzpicture}
\end{center}
\caption{The check gadget.}\label{fig_check}
\end{figure}

The check gadget is illustrated in Figure \ref{fig_check}, whose internal structure is almost identical to the merge gadget. The colors of the neighbors of the two nodes $a$ and $b$ of the check gadget are the same as the merge gadget. Like the merge gadget, the check gadget also has two incoming path gadgets and one out-going path gadget.  The only difference is the position at which the out-going path is attached to the check gadget. This slight difference makes the check gadget behave differently. If the players come from the top, the nodes $T$, $a$, and $B$ get colored in order, and the players leave from the bottom. If the players come from the right, then the game ends at either node $b$ or node $a$, depending on whether the gadget has been traversed before. If node $a$ has been colored in a previous traverse of this check gadget, then the player to color node $b$ creates a rainbow triangle and loses. If node $a$ is uncolored, then node $b$ can be colored green, and the player to color node $a$ loses as he has to create a rainbow triangle, since the distance between node $b$ and $T,B$ is 3.

When two path gadgets cross each other, we need the crossover gadget to keep the one path gadget independent of the other. The crossover gadget is symbolically represented on the left of Figure \ref{fig_crossover}, and it is constructed by one merge gadget, two check gadgets, and three switch gadgets on the right of Figure \ref{fig_crossover}. Note that the lengths of the violet path gadgets are even. If the players come from the red path gadget on the top left, they follow the arrows of the black paths to reach the switch gadget $a$. Because each violet path gadget has an even length, so if one player makes the choice in switch gadget $a$, then it is always the other player to make the choice at switch gadgets $b$ or $c$. Without loss of generality, we assume the hero makes the choice at switch $a$ and the adversary makes the choice at switch gadgets $b$ or $c$. As a result, in order to avoid losing immediately, the hero should choose the path gadget leading to switch gadget $c$. Suppose that his choice leads to switch gadget $b$, then his opponent can choose to go back to the check gadget $x$ at switch gadget $b$. By the fact that the violet path is even and the check gadget $x$ has been traversed, we know that the hero loses. So the hero has to choose the path to switch gadget $c$. This time the adversary would avoid choosing the path leading to the check gadget $y$, because the check gadget $y$ has not been traversed and the adversary will lose. Therefore, the adversary chooses the red path gadget at the bottom right and goes out. In summary, if the players come from the red path gadget, they also leave from the red path gadget. For the same reason, if the players come from the blue path gadget on the top right, they must leave from the blue path gadget at the bottom left.


\begin{figure}[H]
\begin{center}
\begin{tikzpicture}

\node (xing) at (-6,-2) [rectangle, rounded corners, draw] {crossover};
\draw [->, line width=5, color=red] (-8,-2)--(xing)--(-4,-2);
\draw [->, line width=5, color=blue] (-6,-0.5)--(xing)--(-6,-3.5);

\node (c1) at (0,0) [rectangle, rounded corners, draw] {check $x$};
\node (c2) at (4,0) [rectangle, rounded corners, draw] {check $y$};

\node (m) at (2,-1) [rectangle, rounded corners, draw] {merge};

\node (sa) at (2,-3) [rectangle, rounded corners, draw] {switch $a$};

\node (sb1) at (0,-4) [rectangle, rounded corners, draw] {switch $b$};
\node (sb2) at (4,-4) [rectangle, rounded corners, draw] {switch $c$};

\draw [->, line width=5] (c1)--(0,-1)--(m);
\draw [->, line width=5] (c2)--(4,-1)--(m);

\draw [->, line width=5] (m)--(sa);

\draw [->, line width=5, color=violet] (sa)--(0,-3)--(sb1);
\draw [->, line width=5, color=violet] (sa)--(4,-3)--(sb2);

\draw [->, line width=5, color=red] (0,1.2)--(c1);
\draw [->, line width=5, color=red] (sb2)--(4,-5.2);

\draw [->, line width=5, color=blue] (4,1.2)--(c2);
\draw [->, line width=5, color=blue] (sb1)--(0,-5.2);

\draw [->, line width=5, color=violet] (sb1)--(-2,-4)--(-2,0)--(c1);
\draw [->, line width=5, color=violet] (sb2)--(6,-4)--(6,0)--(c2);

\end{tikzpicture}
\end{center}
\caption{The crossover gadget (left) and its implementation (right).}\label{fig_crossover}
\end{figure}
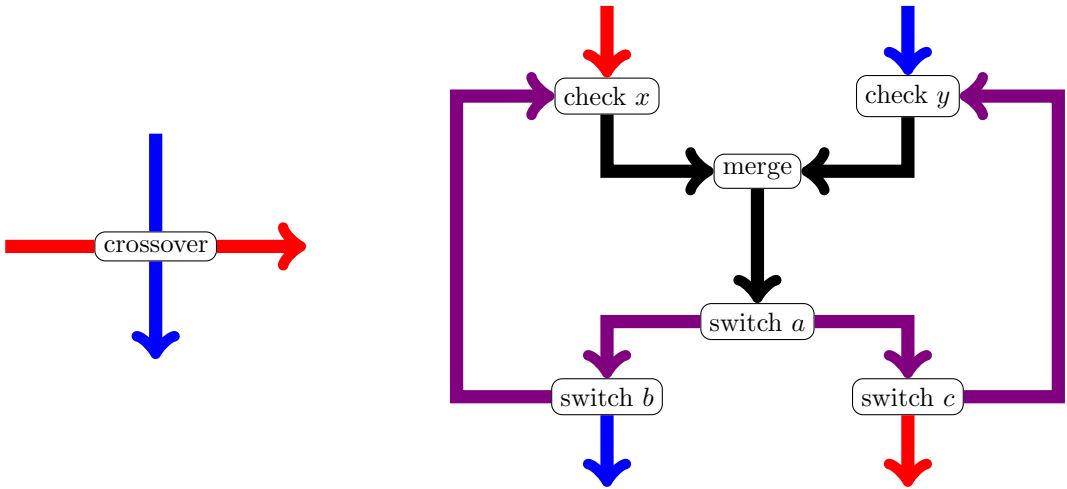

Note also that by the above described mechanism of the crossover gadget, it can be traversed at most once. This is all we need, because in the overall structure illustrated in Figure \ref{fig_structure}, we only need the crossover gadgets near the end of the game. When a literal ($x_i$ or $\lnot x_i$) has been chosen from a clause, the players follow the last path gadgets to go back the the check gadget. It is evidently that a path leading to the check gadget for one literal only intersects with the paths leading to other literals, so each crossover gadget will be traversed at most once during a game play.

Now, we put all the gadgets together. Given a quantified Boolean formula
$$\phi=Q_1 x_1 Q_2 x_2 \cdots Q_n x_n [\psi],$$
where each $Q_i\ (1\leq i \leq n)$ is quantifier (either $\forall$ or $\exists$) and $\psi=c_1 \wedge \dots \wedge c_m$, we can now construct a legal \textsc{Atropos}-$2$ state based on the overall structure (Figure \ref{fig_structure}) with all the gadgets we introduced above. As we have mentioned at the beginning of the proof, we assume the current player of the legal \textsc{Atropos}-$2$ state is the hero.

In the first part of the game play, for each variable $x_i$, there is a switch gadget $Q_i$ for that variable. The players can choose exactly one out-going path gadget from the switch gadget $Q_i$, which simulates the assignment of true or false of $x_i$. We can set the parity (odd or even) of the length of the path gadgets properly so that if the quantifier $Q_i=\forall$, the adversary makes the choice for the out-going path of switch gadget $Q_i$, and if the quantifier $Q_i=\exists$, the hero makes the choice for the out-going path of switch gadget $Q_i$. So for each variable $x_i$, either the hero or the adversary will decide whether to go through a check gadget for $x_i$ or a check gadget for $\lnot x_i$, depending on the choice at switch gadget $Q_i$. Going through the check gadget $x_i$ sets $x_i$ to true, while going through the check gadget $\lnot x_i$ sets $x_i$ to false (hence $\lnot x_i$ is true).

After the assignment of the truth value of every variable in the first part, we come to the second part of the game play. Again, by setting the length of the path gadgets properly, we let the adversary make the choice for the first multi-switch gadget $\psi$ in the second part. There are $m$ out-going path gadgets for multi-switch gadget $\psi$, each leads to a second multi-switch gadget that corresponds to a clause $c_i\ (1\leq i\leq m)$ of the formula $\psi = c_1\wedge\dots \wedge c_m$. The length of the path gadget connecting the first multi-switch gadget $\psi$ and the second multi-switch gadget $c_i$ is even so that it is the hero to make a choice for the second multi-switch gadget. Without loss of generality, we assume the adversary selects the path gadget corresponding to $c_1=x_1 \vee \lnot x_2 \vee \lnot x_n$. In this case, the multi-switch gadget $c_1$ has $3$ out-going path gadgets, connecting to rounded-corner rectangles labelled $x_1$, $\lnot x_2$ and $\lnot x_n$ (see Figure~\ref{fig_structure}), respectively. These three rounded-corner rectangles do not represent any special functional gadgets, and we can just think of each of them as an uncolored node on the path gadget. They are labelled with rounded-corner rectangles so that the overall structure can be seen more clearly. All other multi-switch gadgets $c_j\ (j\neq 1)$ work in the same way as multi-switch gadget $c_1$, except that they may have out-going paths for different literals. Finally, the path gadgets of the same literal ($x_i$ or $\lnot x_i$ from different clauses) merge into one path gadget and go back to the unique check gadget for that literal. By setting the length of the path gadgets properly, we make sure that it is the adversary's turn to color the node labelled $b$ in all the check gadgets for the literals (see Figure~\ref{fig_check}). In short, the adversary chooses a clause, the hero chooses a literal from that clause, and they check the truth value of this literal against the assignment in the first part of the game.

To complete the reduction, we show that the hero (i.e. the current player) has a winning strategy in the legal \textsc{Atropos}-$2$ state we constructed if and only if the quantified Boolean formula $\phi$ is true. If $\phi$ is true, then no matter how the adversary assigns his variables, the hero can assign the rest of the variables such that $\psi=c_1\wedge \dots \wedge c_m$ is true. So $ c_i$ is true for all $1\leq i \leq m$. Therefore, no matter which clause gadget $c_j$ the adversary chooses, there is at least one literal of $c_j$ that is true under the assignment in the first part of the game. The hero then chooses the path gadget heads back to the check gadget for this literal. As the adversary has to color the node $b$ of this check gadget, it creates a rainbow triangle as this check gadget has been traversed before (recall that going through a check gadget in the first part of the game is equivalent to set that the literal of that check gadget true). So the hero wins.

On the other hand, if $\phi$ is false, then no matter how the hero does in assigning the truth value on his part, the adversary can assign the other variables such that $\psi=c_1\wedge \dots \wedge c_m$ is false. So the adversary can choose a clause $c_j$ such that $c_j$ is false. Hence all literals of $c_j$ are false. Then no matter which literal of $c_j$ the hero chooses, it leads to a check gadget which has not been traversed before. So the adversary can color the node $b$ of the check gadget, and the hero colors the node $a$ and creates a rainbow triangle. The adversary wins. This completes the proof of the \textsf{PSPACE}-completeness of \textsc{Atropos}-$2$. 

For the proof of the cases $k\geq 3$, we need to modify the path gadgets so that the distance between two consecutive uncolored nodes is either $1$ or $k$ (allowing adjacent uncolored nodes is handy when we need to adjust the parity of the length of the path gadget). In addition, we should modify the switch gadgets, merge gadgets, and check gadgets to guarantee that the distance between uncolored nodes from different paths is at least $k+1$. 
\hfill$\square$\end{pf}

\section{Conclusion}\label{sec_con}

In this paper, we answer the question asked by Burke and Teng on the computational complexity of \textsc{Atropos}-$k$. It is proved that \textsc{Atropos}-$k$ is PSPACE-complete for any fixed $k\geq 2$. It remains open to decide the computational complexity of \textsc{Atropos}-$\infty$. Since players could color any node at all times in \textsc{Atropos}-$\infty$, it seems that our method can not be applied in this case. 


\section*{References}
\bibliography{atropos_bibfile}

\begin{thebibliography}{10}
\expandafter\ifx\csname url\endcsname\relax
  \def\url#1{\texttt{#1}}\fi
\expandafter\ifx\csname urlprefix\endcsname\relax\def\urlprefix{URL }\fi
\expandafter\ifx\csname href\endcsname\relax
  \def\href#1#2{#2} \def\path#1{#1}\fi

\bibitem{fraenkel1981computing}
A.~S. Fraenkel, D.~Lichtenstein, {Computing a perfect strategy for $n\times n$ {Chess} requires time exponential in $n$}, Journal of Combinatorial Theory, Series A 31~(2) (1981) 199--214.

\bibitem{robson1984checker}
J.~M. Robson, {$N$} by {$N$} {Checkers} is \textsf{EXPTIME} complete, SIAM Journal on Computing 13~(2) (1984) 252--267.

\bibitem{ls80}
D.~Lichtenstein, M.~Sipser, Go is polynomial-space hard, Journal of the ACM 27~(2) (1980) 393–401.

\bibitem{robson1983go}
J.~M. Robson, The complexity of {Go}, in: R.~E.~A. Mason (Ed.), Information Processing 83, Proceedings of the {IFIP} 9th World Computer Congress, Paris, France, September 19-23, 1983, North-Holland/IFIP, 1983, pp. 413--417.

\bibitem{reisch1980gobang}
S.~Reisch, Gobang is \textsf{PSPACE}-complete, Acta Informatica 13 (1980) 59--66.

\bibitem{reisch1981hex}
S.~Reisch, Hex is \textsf{PSPACE}-complete, Acta Informatica 15 (1981) 167--191.

\bibitem{hearn2009games}
R.~A. Hearn, E.~D. Demaine, Games, Puzzles, and Computation, AK Peters/CRC, 2009.

\bibitem{siegel2013combinatorial}
A.~N. Siegel, Combinatorial Game Theory, Vol. 146 of Graduate Studies in Mathematics, American Mathematical Society, 2013.

\bibitem{bt08}
K.~W. Burke, S.-H. Teng, Atropos: A \textsf{PSPACE}-complete {S}perner triangle game, Internet Mathematics 5~(4) (2008) 477--492.

\bibitem{s12}
M.~Sipser, Introduction to the Theory of Computation, 3rd Edition, Cengage Learning, 2012.

\bibitem{iwata1994othello}
S.~Iwata, T.~Kasai, The {Othello} game on an $n \times n$ board is \textsf{PSPACE}-complete, Theoretical Computer Science 123~(2) (1994) 329--340.

\bibitem{cracsmaru2001ladders}
M.~Cr{\^a}{\c{s}}maru, J.~Tromp, Ladders are \textsf{PSPACE}-complete, in: T.~Marsland, I.~Frank (Eds.), Computers and Games, Vol. 2063 of Lecture Notes in Computer Science, Springer, Berlin, Heidelberg, 2001, pp. 241--249.

\bibitem{zhang19}
Z.~Zhang, A note on computational complexity of kill-all {Go} (2019).
\newblock \href {http://arxiv.org/abs/1911.11405} {\path{arXiv:1911.11405}}.

\end{thebibliography}

\end{document}